%% file: full.tex
\input{preamble.tex}

\begin{document}
\bstctlcite{IEEEexample:BSTcontrol}

\title{Pre-Training on Software Engineering Texts:\\Effects on Domain Adaptation and General-Language Understanding

}

 \author{\IEEEauthorblockN{1\textsuperscript{st} Fabian C. Peña \orcidlink{0009-0008-2249-7990}}
 \IEEEauthorblockA{\textit{Faculty of Computer Science and Mathematics} \\
 \textit{University of Passau}\\
 Passau, Germany \\
 fabiancamilo.penalozano@uni-passau.de}
 \and
 \IEEEauthorblockN{2\textsuperscript{nd} Steffen Herbold \orcidlink{0000-0001-9765-2803}}
 \IEEEauthorblockA{\textit{Faculty of Computer Science nd Mathematics} \\
 \textit{University of Passau}\\
 Passau, Germany \\
 steffen.herbold@uni-passau.de}
}


\maketitle

\begin{abstract}
  Generalist and code-focused Language Models (LMs) are increasingly applied to software engineering (SE), yet whether they are optimized for understanding SE textual artifacts (e.g., issues, commit messages, developer discussions) remains unclear, as most evidence comes from code-focused benchmarks. We study how to adapt encoder and decoder LMs to SE text, comparing continual pre-training (CPT) against pre-training from scratch (PTS) on a new SE corpus, and evaluating both domain adaptation (SELU) and general-language understanding (SuperGLUE). To keep the comparisons fair, we control pre-training under constant-token and compute-matched budgets. We find that across families and sizes, reusing an existing LM dominates training a domain-native one from scratch: CPT yields small and mostly inconclusive domain gains while leaving general-language understanding essentially unchanged, whereas PTS pays a large and usually decisive penalty on both axes and becomes competitive only for small LMs under a token-rich budget. We distill these results into practical guidance for adapting LMs to SE text and release our corpus and pre-trained LMs in our replication kit~\cite{replication-kit}.
\end{abstract}


\section{Introduction}

In recent years, Language Models (LM) have been adopted in a growing array of professional tasks, with particularly visible use in software engineering (SE) and other knowledge-work activities~\cite{anthropic-index}. Although empirical studies report substantial productivity gains in some controlled settings, these effects are highly heterogeneous and depend on task structure and worker expertise~\cite{developer-productivity,productivity}. Frontier-LM development is evolving at a rapid pace, with successive releases introducing improvements in capabilities, efficiency, and alignment~\cite{openai-gpt55,anthropic-opus47,kimi-kimik26}; however, evaluating these improvements usually depends on narrow benchmarks which lack real-world representativeness. Meanwhile, open questions remain regarding reliability, robustness~\cite{reliability,hallucinations}, and the extent to which apparent general capabilities transfer to specialized domains. In SE, for example, popular code-focused benchmarks such as SWE-bench~\cite{swebench} and Terminal-bench~\cite{terminalbench} evaluate LMs on issue resolution and CLI handling, respectively, leaving other aspects of the SE practice in the dark.

Only a recent benchmark, SELU~\cite{selu}, enables evaluating language understanding capabilities through the broader SE domain with a cornerstone in SE textual artifacts specific to code-adjacent metadata, pure non-code settings, and developer communications. By running SELU on a range of LMs, the authors found that those that might be considered outdated (e.g., GPT-2~\cite{gpt2}) may outperform larger, more recent and even code-focused LMs (e.g., Llama 3.2~\cite{llama32} and StarCoder2~\cite{starcoder2}). We hypothesize that this improvement stems from optimizing LMs for SE textual artifacts from a foundational stage (i.e., pre-training), feeding them high-quality data aligned with this purpose.

Based on this hypothesis, we systematically study domain adaptation by applying continual pre-training (CPT)~\cite{domain-adaptation1,domain-adaptation2,domain-adaptation3} to a range of open-source LMs on a new corpus of SE text collected from sources such as GitHub, Stack Overflow, Jira, and arXiv. Similarly to CPT, we also pre-train LMs from scratch (PTS), i.e., not starting from an existing LM but from a new one with randomly initialized weights. Throughout, domain adaptation refers to adapting LMs to SE textual artifacts, the natural language developers produce around code (e.g., issues, commit messages, developer discussions), rather than source code itself. During pre-training, we control the token and compute budgets to account for scaling laws~\cite{scaling-laws, chinchilla}. Our results show that CPT can slightly improve domain adaptation measured with SELU, with a possible small loss in general-language understanding measured with SuperGLUE~\cite{superglue}. Following the same evaluation protocol, PTS proves to be inferior. Still, the advantage of CPT is small, especially for the most recent models we study, indicating that robust general-domain training may be sufficient for SE textual artifact tasks without domain adaptation.

\section{Background and Related Work}
\label{s:background}

Today, CPT is a widely used route for domain adaptation, where an existing LM (checkpoint) is shifted toward a new unlabeled data distribution. Initial work on domain-adaptive and task-adaptive pre-training showed gains across several low- and high-resource domains~\cite{domain-adaptation1}. Subsequent work reframed this idea as lifelong or continual pre-training under sequential distribution shift and studied replay, distillation, and masking-based strategies to retain previously acquired knowledge~\cite{domain-adaptation2,domain-adaptation3}. More specifically, FinPythia showed that CPT on finance data can improve domain-specific performance and, with selected domain subsets, preserve open-domain benchmark scores~\cite{finpythia}; and TiC-LM studied time-continual pre-training on general web data and domains such as Wikipedia, Stack Exchange, and code documentation, showing that replay can approach periodic retraining from scratch at lower compute cost, with replay being more important on generic web data than on narrower domains~\cite{ticlm}. These studies establish CPT as a strong baseline for domain adaptation, which may also erode previously acquired capabilities~\cite{forgetting}. In contrast, other works also suggest when PTS may be worthwhile. For example, PubMedBERT demonstrated that with abundant unlabeled domain-specific data, PTS can outperform CPT from a generalist LM~\cite{pubmedbert}.

Our budget-aware experimental design is motivated by leading scaling-laws research that showed training loss follows regular power-law relationships with LM size, number of tokens, and compute spent for training~\cite{scaling-laws}. Under a compute-optimal training, the number of non-embedding parameters and the number of training tokens should scale much more closely together to avoid undertraining~\cite{chinchilla}. This budget allocation framework becomes especially important for comparing pre-training experiments, since otherwise gains may reflect budget differences rather than adaptation quality. Recent literature suggests that encoders are even more data-hungry than decoders~\cite{compute-optimal-encoders}.

Specific to the application of LMs to SE, studies are skewed toward code-focused tasks, with individual efforts spread across requirements classification and elicitation, traceability, issue management, documentation, logging, and software-management tasks~\cite{llm4se-survey}. Prior work on training SE-specific LMs already hints that specialized pre-training can help on textual artifact tasks. BERTOverflow improved Stack Overflow named entity recognition over off-the-shelf BERT~\cite{bertoverflow}. The authors of seBERT found that pre-training on SE text is valuable when tasks depend on SE context (e.g., domain terminology), while generalist LMs remain sufficient for general-language understanding, also within the SE domain~\cite{validity-se}. SOBert demonstrated that medium, domain-specific LMs can outperform larger generalist and code-focused LMs on several Stack Overflow text understanding tasks~\cite{sobert}. These three works remain localized in Stack Overflow or narrow task settings, so it is still unclear whether a broader SE corpus may give grounds for PTS and whether potential gains transfer across diverse non-code and code-adjacent tasks. The recently introduced SELU benchmark helps address this opportunity by enabling the evaluation of LM understanding, to the best of our knowledge, across a broader range of SE textual artifact tasks~\cite{selu}, making it a particularly suitable domain-specific testbed for our setting.

Based on this, we observe a gap in applying pre-training on SE text data, mainly for decoder LMs, and evaluating the trade-off between potential domain adaptation gains and general-language understanding losses in token and compute budget controlled settings.

\section{Research Questions}
\label{s:rqs}

In the context of applying CPT and PTS to a range of open-source LMs on a new corpus of SE text and evaluating them for both domain adaptation and general-language understanding, we study the following research question:

\begin{description}
    \item[\textbf{RQ}] How do different schemes (CPT and PTS) for adapting LMs to the SE domain compare to each other with respect to token budget, family, and size?
\end{description}

Within our research, we consider this question from four angles: i) does CPT on domain-specific data improve the understanding of SE textual artifacts? ii) does this negatively affect general-language retention? iii) how does PTS on SE data compare to CPT for the understanding of SE textual artifacts? and iv) how much general-language understanding emerges when applying PTS on SE data alone?

\section{Experimental Setup}
\label{s:setup}

In this section we define the methodological elements our experiments are based on: our new corpus of SE text, the LMs selected for domain adaptation, the analytical method to define the constant-token and compute-matched budgets to run the experiments, the pre-training details for both CPT and PTS schemes, and the evaluation protocol.

\subsection{Our SE Corpus}
\label{ss:corpus}

A central contribution of this work is a new, large-scale corpus of SE text assembled specifically for pre-training LMs on the language of the SE practice rather than on source code. We deliberately target natural-language artifacts that developers and researchers produce around code (e.g., issues, commit messages, developer discussions, papers) and mask code blocks throughout, so adaptation is driven by SE textual patterns alone; this distinguishes our work from code-focused adaptation efforts that dominate the literature~\cite{codellama,starcoder2}.

\subsubsection{Source Composition}

To build our SE corpus, we collect publicly available data from GitHub, Stack Overflow, Jira, and arXiv, still recognizing the lack of representativeness in terms of the broader SE practice (e.g., requirements analysis). Table~\ref{tbl:se-corpus-final} shows the main characteristics per dataset.

\begin{table*}[ht]
\centering
\caption{Composition of our SE corpus. File size is in parquet format and tokens based on the WordPiece tokenizer.}
\label{tbl:se-corpus-final}
\begin{tabular}{l cccc cc cccc} 
\toprule
\multirow{2}{*}{\textbf{Source / Dataset}} &
\multirow{2}{*}{\textbf{Time window}} &
\multirow{2}{*}{\textbf{File size}} & 
\multicolumn{2}{c}{\textbf{Before preparation}} &
\multicolumn{2}{c}{\textbf{After preparation}} &
\multicolumn{4}{c}{\textbf{\textbf{\textbf{\textbf{Document length statistics}}}}} \\ 
\cline{4-11}
&&&
\textbf{\# Docs.} &
\textbf{\# Tokens} &
\textbf{\# Docs.} &
\textbf{\# Tokens} & \textbf{min.} & \textbf{avg.} & \textbf{max.} & \textbf{[p10, p50, p90]} \\ 
\hline
GitHub & 2015-2025 & 58.1GB & 108.1M & 64.6B & 71.3M & 11.2B & 1 & 157 & 214.2K & [11, 71, 280] \\
Stack Overflow & 2008-2025 & 31.9GB & 58M & 25.4B & 57M & 6.8B & 1 & 120 & 20.5K & [26, 91, 236] \\
Jira & 2010-2022 & 809MB & 2.2M & 789.7M & 2M & 423.9M & 1 & 213 & 2.8M & [24, 86, 313] \\
arXiv & 1998-2025 & 14.7MB & 20.9K & 5.4M & 20.8K & 5.3M & 15 & 256 & 699 & [146, 255, 370] \\ 
\hline
 &  &  & \textbf{168.3M} & \textbf{90.8B} & \textbf{130.3M} & \textbf{18.5B} &  &  &  &  \\
\bottomrule
\end{tabular}
\end{table*}

GitHub is the largest forge used by 180M+ developers, having 420M+ repositories~\cite{github-about}. It centralizes source code, issues, and pull requests. We download from the GitHub Archive~\cite{github-archive} 590.7M issues and pull requests (title + body) and only keep those with any activity other than posting by looking at comments and reviews giving us 108.1M documents.
    
Stack Overflow has been for many years the de facto meeting place for developers to answer queries about programming, among other topics; and recent surveys show that $\sim$25\% of those surveyed use it at least once a day~\cite{stackoverflow-survey-2025}, although with a downward trend compared to previous years~\cite{stackoverflow-survey-2024,stackoverflow-survey-2023}. We download from the Internet Archive~\cite{internet-archive} 60.5M Stack Overflow and Software Engineering Stack Exchange posts (title + body) and keep those that do not have a negative voting (i.e., any post without votes is also included).
    
Jira is a collaborative work management and code documentation tool, just recently surpassed by GitHub in popularity~\cite{stackoverflow-survey-2024}. It is decentralized, i.e., every project hosts its own instance, which is why relevant open-source projects (e.g., Apache and MariaDB) prefer it for managing community issues. We rely on a previous effort~\cite{jira-dataset} to collect a Jira dataset from 16 organizations with a total of 1,822 projects and 2.7M issues (summary + description). Similarly to GitHub, we keep issues with any activity other than posting by looking at comments.
    
ArXiv concentrates the major research effort in SE as well as in many other domains. The official dataset available on Kaggle~\cite{arxiv-dataset} is regularly updated with metadata from the latest submissions. We target those in the cs.SE category for a total of 20.9K papers (titles + abstracts).

Despite their volume, we discard documents such as commit messages and comments, since these tend to be much shorter than issues or posts, and their quality is harder to estimate.

\subsubsection{Data Preparation}

To derive a high-quality SE corpus aligned with our research objectives, we apply standard yet comprehensive data preparation techniques to our initial SE corpus as described below. 

We implement different preprocessing pipelines tailored to each dataset with some common steps including: (i) normalizing white spaces, (ii) removing any kind of markup, and (iii) masking by some common regex patterns, such as URLs, hashes, user mentions, and code blocks. While arXiv is mostly plain text, GitHub documents come in markdown, Stack Overflow in HTML, and Jira uses its own markup format. The rationale behind masking code blocks is that we decide to evaluate LMs on their capability to capture SE textual artifact patterns, based solely on the context that natural language can provide. This differentiates the results of this work from code-focused LM adaptation studies found predominantly in the literature~\cite{codellama,starcoder2}. To isolate cross-language effects, we remove non-English documents using a language identification model based on fastText~\cite{fasttext1,fasttext12}. After these, we observe drops in the number of documents, from 168.3M to 158.5M ($\sim$5.8\%), and in the number of tokens, from 90.8B to 30.9B ($\sim$66\%). 

We apply deduplication and decontamination processes using the MinHash algorithm~\cite{minhash} implemented in the DataTrove library~\cite{datatrove}. More details can be found in the replication kit~\cite{replication-kit}. While deduplication is performed independently for each dataset in our corpus, decontamination is performed against SELU~\cite{selu}, the benchmark we use to evaluate domain adaptation (see Subsection~\ref{ss:evaluation}) to prevent information leaks. We observe additional drops in the number of documents, from 158.5M to 130.3M ($\sim$17.8\%), and in the number of tokens, from 30.9B to 18.5B ($\sim$40.1\%). Decontamination against SELU has a relatively high contribution to this drop (516.2K documents) considering that the total number of instances in this benchmark is 1.1M. This is expected since most of the SELU tasks were built from the same sources as our corpus.

We split every dataset before starting any pre-training experiment, reserving 5\% of the documents for validation. The reason for this relatively low proportion is that we are more interested in evaluating LMs on downstream tasks (i.e., domain adaptation and general-language understanding) rather than on proxy learning objectives (see Subsection~\ref{ss:evaluation}). The validation split is, therefore, used only to monitor training.

\subsection{LMs Selected for Domain Adaptation}
\label{ss:lms-pretraining}

In our study, two pre-training schemes for domain adaptation compete: continual pre-training (CPT)~\cite{domain-adaptation1,domain-adaptation2,domain-adaptation3} and pre-training from scratch (PTS). Both use the same LM architectures, learning objectives, and optimization algorithms. What makes them different is the initial state of the LM (checkpoint). In CPT, the initial LM parameters or weights follow a distribution that represents the knowledge already acquired in a previous pre-training stage. In PTS these are randomly initialized. In both schemes, these parameters are updated based on a learning objective: for encoder LMs, predicting a randomly masked in-between token (Masked Language Modeling, MLM), and for decoder LMs, predicting the next token (Causal Language Modeling, CLM).

We select a variety of encoder and decoder open-source LMs with numbers of parameters ranging from 108M to 7.2B. The complete list can be found in Table~\ref{tbl:llms-configs}. Notice that for comparison purposes, we use the name, e.g., GPT-2 small, to reference the official checkpoint (i.e., the initial state of the LM) used in CPT, but also a new LM that is PTS and instantiated using an equivalent configuration. For CPT, we use the same tokenizer published alongside the official checkpoint. For PTS, we train new tokenizers on our SE corpus from scratch, matching the algorithm and vocabulary size for each LM family. Previous work has shown that such SE-tokenizers are distinguishable from those used for generalist LMs~\cite{validity-se}. Official checkpoints, configurations, and tokenizers are taken from Hugging Face~\cite{huggingface_hub}.


\begin{table*}[ht]
\centering
\caption{LMs selected for domain adaptation and their corresponding pre-training configuration. First compartment: Generalist LMs. Second compartment: Code-focused LMs. Token budgets separated by comma produce different checkpoints. Batch size is given per GPU. CodeLlama 7B and StarCoder2 7B run with gradient checkpointing.}
\label{tbl:llms-configs}
\begin{tabular}{clccccccc} 
\toprule
& \multicolumn{1}{c}{\textbf{LM family}} & \textbf{Architecture} & \textbf{\# Params.} & \textbf{Scheme} & \textbf{Block sz.} & \begin{tabular}[c]{@{}c@{}}\textbf{Token budget (acc.)}\end{tabular} & \begin{tabular}[c]{@{}c@{}}\textbf{Batch sz.}\end{tabular} & \begin{tabular}[c]{@{}c@{}}\textbf{Grad. acc. steps}\end{tabular} \\ 
\hline
\multirow{2}{*}{\cite{bert}} & BERT base & \multirow{2}{*}{encoder} & 108M & \multirow{2}{*}{CPT / PTS} & \multirow{2}{*}{512} & 3.3B, 11.7B & \multirow{2}{*}{16} & \multirow{2}{*}{1} \\
 & BERT large & & 334M & &  & 3.3B &  &  \\ 
\hdashline[1pt/1pt]
\multirow{2}{*}{\cite{roberta}} & RoBERTa base & \multirow{2}{*}{encoder} & 125M & \multirow{2}{*}{CPT / PTS} & \multirow{2}{*}{512} & 3.3B, 11.7B & \multirow{2}{*}{16} & \multirow{2}{*}{1} \\
 & RoBERTa large & & 355M & &  & 3.3B &  &  \\  
\hdashline[1pt/1pt]
\multirow{5}{*}{\cite{modernbert}} & ModernBERT base & \multirow{2}{*}{encoder} & 150M & \multirow{2}{*}{CPT} & \multirow{2}{*}{8K} & 3.3B, 9.0B & \multirow{2}{*}{4} & \multirow{2}{*}{4} \\
 & ModernBERT large & & 396M & &  & 2.9B, 3.3B &  &  \\ 
\cdashline{3-9}[1pt/1pt]
 & ModernBERT base & \multirow{3}{*}{encoder} & 150M & PTS & \begin{tabular}[c]{@{}c@{}}1K\\8K\end{tabular} & \begin{tabular}[c]{@{}c@{}}3.3B, 7.7B\\9.0B\end{tabular} & \begin{tabular}[c]{@{}c@{}}16\\4\end{tabular} & \begin{tabular}[c]{@{}c@{}}1\\4\end{tabular} \\ 
\cdashline{4-9}[1pt/1pt]
 & ModernBERT large &  & 396M & PTS & \begin{tabular}[c]{@{}c@{}}1K\\8K\end{tabular} & \begin{tabular}[c]{@{}c@{}}2.5B\\2.9B, 3.3B\end{tabular} & \begin{tabular}[c]{@{}c@{}}16\\4\end{tabular} & \begin{tabular}[c]{@{}c@{}}1\\4\end{tabular} \\ 
\hdashline[1pt/1pt]
\multirow{4}{*}{\cite{gpt2}} & GPT-2 small & \multirow{4}{*}{decoder} & 124M & \multirow{4}{*}{CPT / PTS} & \multirow{4}{*}{1K} & 3.3B, 11.8B & \multirow{3}{*}{16} & \multirow{3}{*}{1} \\
 & GPT-2 medium &  & 355M &  &  & 3.3B &  &  \\
 & GPT-2 large &  & 774M &  &  & 1.4B, 3.3B &  &  \\
\cdashline{7-9}[1pt/1pt]
 & GPT-2 xl &  & 1.6B &  &  & 678.5M, 3.3B & 8 & 2 \\ 
\hdashline[1pt/1pt]
\multirow{5}{*}{\cite{llama32}} & Llama 3.2 1B & \multirow{2}{*}{decoder} & 1.2B & \multirow{2}{*}{CPT} & \multirow{2}{*}{8K} & 1.0B, 3.3B & 2 & 8 \\
 &Llama 3.2 3B &  & 3.2B &  &  & 355.2M, 3.3B & 1 & 16 \\ 
\cdashline{3-9}[1pt/1pt]
 & Llama 3.2 1B & \multirow{3}{*}{decoder} & 1.2B & PTS & \begin{tabular}[c]{@{}c@{}}1K\\8K\end{tabular} & \begin{tabular}[c]{@{}c@{}}874.5M\\1.0B, 3.3B\end{tabular} & \begin{tabular}[c]{@{}c@{}}16\\2\end{tabular} & \begin{tabular}[c]{@{}c@{}}1\\8\end{tabular} \\ 
\cdashline{4-9}[1pt/1pt]
 & Llama 3.2 3B &  & 3.2B & PTS & \begin{tabular}[c]{@{}c@{}}1K\\8K\end{tabular} & \begin{tabular}[c]{@{}c@{}}301.9M\\355.2M, 3.3B\end{tabular} & \begin{tabular}[c]{@{}c@{}}8\\1\end{tabular} & \begin{tabular}[c]{@{}c@{}}2\\16\end{tabular} \\ 
\hline
\cite{codebert} & CodeBERT base & encoder & 125M & CPT & 512 & 3.3B, 11.7B & 16 & 1 \\ 
\hdashline[1pt/1pt]
\cite{codellama} & CodeLlama 7B & decoder & 6.7B & CPT & 8K & 151.5M, 3.3B & 4 & 4 \\ 
\hdashline[1pt/1pt]
\multirow{2}{*}{\cite{starcoder2}} & StarCoder2 3B & \multirow{2}{*}{decoder} & 3.0B & \multirow{2}{*}{CPT} & \multirow{2}{*}{8K} & 347.7M, 3.3B & 1 & 16 \\
 & StarCoder2 7B &  & 7.2B &  &  & 144.1M, 3.3B & 4 & 4 \\
\bottomrule
\end{tabular}
\end{table*}

In addition to the number of parameters and architecture, the selected LMs differ in terms of the type of data that was used to pre-train the official checkpoint. We have (i) generalist LMs that were pre-trained mostly on natural language data without a particular focus on a knowledge domain; and (ii) code-focused LMs that were fed with large amounts of source code data in a variety of programming languages with the expectation that they perform well on code tasks. We exclude code-focused LMs from PTS, as our SE corpus contains no source code and their training requires code as a modality. Larger LMs could not be selected due to the compute required for experimentation.

\subsection{Pre-Training Budgets and Comparison Protocol}
\label{ss:budget-comparison}

We estimate the pre-training effort for each LM following two different protocols: constant-token budget and compute-matched budget. Under a constant-token budget, comparisons isolate the effect of pre-training scheme by holding exposure to our SE corpus fixed. Because we additionally control the effective global batch size and derive the number of training steps directly from the token budget (see Subsection~\ref{ss:pretraining-details}), these comparisons also keep the optimization schedule comparable across LMs. Comparing CPT and PTS under equal token exposure reveals the extent to which inherited general-language priors improve data efficiency. The target constant-token budget is set to 3.3B tokens, matching the number of tokens originally used to pre-train BERT~\cite{bert}.

Under a compute-matched budget, comparisons isolate differences under approximately equal compute cost. In particular, comparing CPT and PTS under matched compute cost reveals how inherited general-language priors improve compute efficiency. To estimate compute costs, we draw motivation from a previous work that established a relationship between the total FLOPS ($C$) required by a LM on a forward and backward pass, its number of non-embedding parameters and the number of tokens ($D$) that are fed on that pass~\cite{scaling-laws, chinchilla}:

\begin{equation}
\label{eq:flops-vs-tokens}
C \approx 6 \cdot N \cdot D
\end{equation}

\begin{figure}[ht]
\centering
\caption{Overview of the budget-alignment protocol. (a) Experiments comparable under the constant-token budget of 3.3B. (b) Experiments comparable under the compute-matched budget of $6.01 \times e^{18}$ FLOPS, where LMs are pre-trained on different number of tokens but at approx. equal compute cost.}
\label{fig:comparisons_under_budget}
\includegraphics[width=0.49\textwidth]{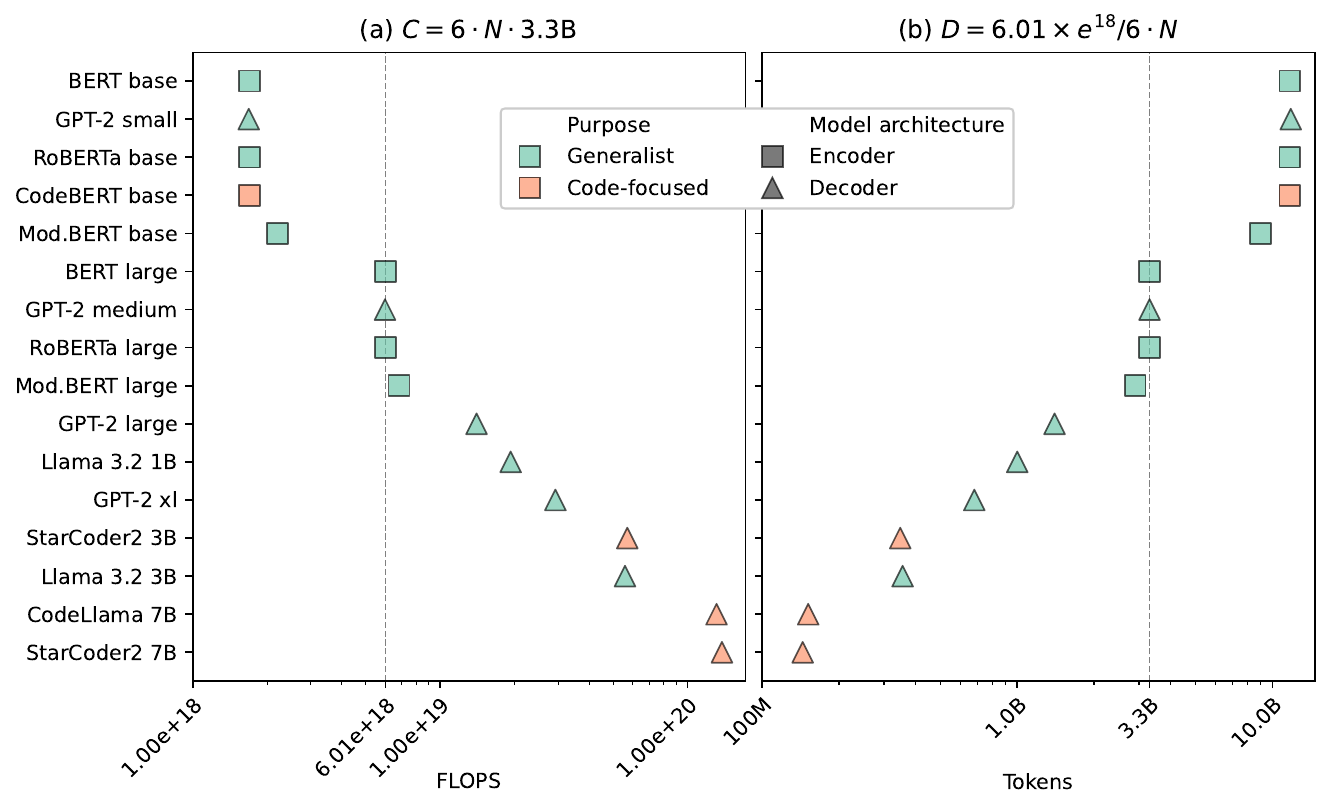}
\end{figure}

We take $6.01 \times e^{18} = 6 \cdot 303.4\text{M} \cdot 3.3\text{B}$ as a reference budget representing an estimate of the compute cost originally spent to pre-train BERT large~\cite{bert}, where $303.4$M is its number of non-embedding parameters. Using this equation, we calculate the number of tokens for each LM in Table~\ref{tbl:llms-configs} that matches this reference budget (see Figure~\ref{fig:comparisons_under_budget}).

Although Equation~\ref{eq:flops-vs-tokens} was mainly derived for decoder LMs, we do not consider differences between architectures and family-specific optimizations for practical purposes (e.g., local and global attention in ModernBERT~\cite{modernbert}). Long sequences during pre-training might also play a role in compute costs~\cite{scaling-laws}, but since most of the documents in our SE corpus are relatively short (see Table~\ref{tbl:se-corpus-final}), we decide not to take full advantage of extended context windows in LMs such as Llama~3.2. See Subsection~\ref{ss:pretraining-details} for more details.

\subsection{Pre-Training Details}
\label{ss:pretraining-details}

We configure CPT and PTS mostly in the same way. Experiment inputs include token budgets (both constant-token and compute-matched, producing intermediate and final checkpoints), block size, batch size per GPU, and gradient accumulation steps. We summarize the experiment configuration for each LM in Table~\ref{tbl:llms-configs} and provide more details below. All experiments run on a server with 8 NVIDIA A100 GPUs and, for supporting LMs up to 7B parameters, we use DeepSpeed (stage 2)~\cite{deepspeed}, which shards gradients and optimizer states across GPUs. We load the LM configurations using Transformers~\cite{transformers} and implement the pre-training logic on Accelerate~\cite{accelerate}.

During training, all datasets in our SE corpus are loaded into memory in streaming mode. We then merge them with an interleaving strategy that takes alternate documents (hereafter referred to as sequences) from each dataset with different sampling probabilities. Since our SE corpus has an unequal composition (i.e., $\sim$60\% of tokens come from GitHub but less than 1\% come from arXiv), we decide on probabilities that roughly exhibit this composition but also with some guarantee of representativeness during sequence packing and data collation later. Specifically, we set a probability of 0.45 for GitHub and Stack Overflow, and 0.05 for Jira and arXiv. Studying the effect of different interleaving probabilities is out of scope in this work due to compute requirements.

We further post-process our data to remove any sequences with fewer than 17 tokens. We then pack (i.e., put together multiple sequences into one, adding a corresponding end-of-sequence token) and slice them to the desired block size. For CPT, this block size is either the full context window supported by the official checkpoint (e.g., 512 for BERT-like LMs or 1K for GPT-2), or a capped value that we define for LMs that support extended context windows (i.e., Llama 3.2, CodeLlama, and StarCoder2). We define this capped block size as 8K, taken from ModernBERT as a reference, being aware that most of the sequences from our SE corpus do not even fill a 512 block size individually (see Table~\ref{tbl:se-corpus-final}). For PTS ModernBERT and Llama 3.2 LMs, inspired by their original pre-training recipes~\cite{modernbert,llama32}, we start pre-training with a block size of 1K for about 85\% of the corresponding compute-matched tokens, then we increase it to 8K. This helps the attention mechanism to first adapt to shorter token co-occurrences before trying to learn long-context patterns. We keep the token-masking probability the same as that used for pre-training the official checkpoints (e.g., a probability of 0.15 for any BERT-like LM~\cite{bert,roberta} and 0.3 for ModernBERT~\cite{modernbert}).

For defining the number of pre-training steps to run an experiment, we take the largest token budget and block size, together with the batch size per GPU and gradient accumulation steps that match a global batch size of 128. Other hyper-parameters that are kept constant across experiments include using AdamW as the optimizer, a weight decay of 0.01, a maximum learning rate of $1 \times e^{-5}$ with a warm-up period equivalent to 5\% of the total steps, a cosine decreasing factor to a minimum learning rate of 10\% of the maximum value, and a global seed of 42. Since our study prioritizes comparisons under equivalent token and compute budgets versus merely optimizing for the learning objective during pre-training, we find these values for global batch size and maximum learning rate to be a good compromise across LMs to progress in the reduction of training and validation losses while effectively using the available GPU memory.

\subsection{Evaluation Protocol}
\label{ss:evaluation}

We evaluate all LMs (the domain-adapted and the official checkpoints) from two different perspectives: SELU~\cite{selu} to evaluate domain adaptation and SuperGLUE~\cite{superglue} to evaluate general-language understanding. SELU is a multi-task, heterogeneous evaluation suite with 22 SE textual artifact tasks. We evaluate our LMs only on the 20 classification and regression tasks that can be supported by both encoder and decoder LMs under fine-tuning. The definitive list of the SELU tasks included in our evaluation protocol is summarized in Table~\ref{tbl:tasks}. Taking the pattern from SELU, we select F1-macro and SMAPE as the evaluation metrics for classification and regression tasks, respectively; then we take the average of these results as the SELU score per LM. To the best of our knowledge, SELU is the most comprehensive benchmark to evaluate adaptation to SE with a particular focus on understanding of textual artifacts. 

CPT typically comes at the cost of degrading capabilities that LMs have previously acquired \cite{forgetting}. Although some mitigation techniques have been proposed in the literature (see Subsection~\ref{s:background}), this remains an open issue in LM research. We do not apply specific mitigation techniques, but rather measure general-language understanding using SuperGLUE~\cite{superglue}. Similar to SELU, this is a multi-task, heterogeneous benchmark with complementary properties: It evaluates language understanding and basic reasoning capabilities, and its tasks were designed to be solvable by most college-educated English speakers without requiring domain-specific knowledge. Accuracy is the main evaluation metric for five of the tasks, and the remaining tasks use complementary metrics to measure different aspects of error: F1-macro, answer-level F1, and Exact Match. We take the average of results as the SuperGLUE score per LM. While for CPT the role of SuperGLUE is to evaluate general-language retention, for PTS, its role is to evaluate general-language emergence from our SE corpus alone, i.e., with no exposure to broad general-domain text. 

\begin{table*}[ht]
\centering
\caption{List of SELU and SuperGLUE tasks to evaluate domain adaptation and general-language understanding. For SELU, all classification tasks use F1-macro and \textit{story\_points} uses SMAPE as evaluation metric. For SuperGLUE, \textit{cb} uses both accuracy and F1-macro, and \textit{multirc} uses Exact Match and answer-level F1. The remaining tasks only use accuracy.}
\label{tbl:tasks}
\begin{tabular}{ccllcc} 
\toprule
\textbf{Benchmark} & \textbf{Task type} & \multicolumn{1}{c}{\textbf{Task ID}} & \multicolumn{1}{c}{\textbf{Task definition}} & \textbf{Instances} & \textbf{Targets} \\ 
\midrule
\multirow{20}{*}{SELU} & \multirow{14}{*}{\rotcell{Binary\\\& multi-class}} & \textit{bug\_issue} & Is the issue reporting a bug? & 38,219 & 2 \\
 &  & \textit{functional\_requirement} & Does the requirement include some functional aspect? & 956 & 2 \\
 &  & \textit{incivility} & Does the text show unnecessary rude behavior? & 1,546 & 2 \\
 &  & \textit{quality\_requirement} & Does the requirement include some quality aspect? & 956 & 2 \\
 &  & \textit{safety\_issue} & Is the issue reporting safety-related concerns? & 1,916 & 2 \\
 &  & \textit{security\_requirement} & Does the requirement include some security aspect? & 510 & 2 \\
 &  & \textit{tone\_bearing} & Does the text have an unnecessarily disrespectful tone? & 6,597 & 2 \\
 &  & \textit{closed\_question} & Which is the reason for closing the question after moderation? & 140,272 & 5 \\
 &  & \textit{commit\_intention} & Is the commit perfecting or correcting the code? & 2,533 & 3 \\
 &  & \textit{issue\_intention} & Which is the intention expressed in the issue? & 6,375 & 7 \\
 &  & \textit{issue\_type} & Is the issue related to a bug, an enhancement or a question? & 803,417 & 3 \\
 &  & \textit{question\_quality} & Is the question of good quality or does it require moderation? & 60,000 & 3 \\
 &  & \textit{review\_type} & Which is the intention expressed in the app review? & 1,390 & 4 \\
 &  & \textit{sentiment} & Which is the sentiment expressed in the text? & 13,144 & 3 \\ 
\cmidrule{2-6}
 & \multirow{5}{*}{\rotcell{Multi-label}} & \textit{comment\_type\_java} & \multirow{3}{*}{Which kind of contents are detailed in the code comment?} & 9,339 & 7 \\
 &  & \textit{comment\_type\_pharo} &  & 1,587 & 7 \\
 &  & \textit{comment\_type\_python} &  & 2,290 & 5 \\
 &  & \textit{review\_aspect} & Which aspects are involved in the API review? & 4,522 & 11 \\
 &  & \textit{smell\_doc} & Which does the API documentation smell like? & 1,000 & 5 \\ 
\cmidrule{2-6}
 & Regression & \textit{story\_points} & What is the effort estimated for the development task? & 23,313 & {[}1-96] \\ 
\midrule
\multirow{7}{*}{SuperGLUE} & \multirow{7}{*}{\rotcell{Binary\\\& multi-class}} & \textit{boolq} & Simple QA: yes, no & 12,697 & 2 \\
 &  & \textit{cb} & Language inference: entailment, contradiction, neutral & 306 & 3 \\
 &  & \textit{copa} & Cause-effect reasoning from two possible choices & 500 & 2 \\
 &  & \textit{multirc} & QA with multiple possible correct answers & 32,091 & 2 \\
 &  & \textit{rte} & Language inference: entailment, not entailment & 2,767 & 2 \\
 &  & \textit{wic} & Word sense desambiguation from two sentences & 6,066 & 2 \\
 &  & \textit{wsc} & Coreference resolution from pronouns and nouns & 658 & 2 \\
\bottomrule
\end{tabular}
\end{table*}

We extend the LMs with trainable components (one pooling operation, one dropout layer, one linear layer, and one activation function). These have to be fitted on each SELU and SuperGLUE task during a fine-tuning stage, resulting in new individual LM checkpoints. For this, we set together a batch size per GPU and a number of gradient accumulation steps that match a global batch size of 16. Combinations range from 4 to 16 for batch size per GPU and from 1 to 4 for gradient accumulation steps, looking to effectively use the available GPU memory. Other hyper-parameters that we keep constant across experiments include using AdamW as the optimizer, a weight decay of 0.01, total steps equivalent to 10 epochs with an early stopping criterion of 3 epochs, a maximum learning rate of $1 \times e^{-5}$ or $1 \times e^{-4}$ (see more details below) with a warm-up period equivalent to 10\% of the total steps, then kept constant, and a global seed of 42. We initially use a maximum learning rate of $1 \times e^{-5}$ everywhere, but identify convergence issues when fine-tuning on datasets below 10K instances. In these cases, we decide to use a maximum learning rate of $1 \times e^{-4}$ instead. We do not report the list of hyper-parameters for all experiments due to space, but they can be found in our replication kit~\cite{replication-kit}.

We report both absolute scores and score deltas. Absolute scores reflect the overall benchmark performance of a LM on SELU or SuperGLUE, while score deltas make performance changes relative to the official checkpoint explicit. Concretely, for each LM, we define $\Delta \text{SELU}$ as the difference between the SELU score of a given LM adapted to our SE corpus and that of its corresponding official checkpoint and analogously define $\Delta \text{SGLUE}$ with respect to SuperGLUE. In CPT, $\Delta \text{SELU}$ and $\Delta \text{SGLUE}$ are interpreted as measures of domain adaptation gain and general-language loss, respectively. In PTS, $\Delta \text{SELU}$ and $\Delta \text{SGLUE}$ measure the acquisition of capability compared to the official checkpoint. Results are reported separately for the constant-token and compute-matched comparison groups defined in Subsection~\ref{ss:budget-comparison}.

Beyond point estimates, we conduct a Bayesian assessment of whether observed differences are significant. For each pair of LMs A and B (A the official checkpoint and B the domain-adapted LM of the same family and size), we form the paired vectors of per task scores across the 20 SELU and 7 SuperGLUE tasks and define a Region of Practical Equivalence (ROPE) of $\pm 0.1 \cdot d$, with $d$ the Cohen's effect size~\cite{cohens-effect} computed from the pooled standard deviation. We then apply the Bayesian signed-rank test~\cite{bayesian-test}, as implemented in Autorank~\cite{autorank}, which returns three posterior probabilities (PP), $P(A>B)$, $P(A=B)$, and $P(B>A)$, for the official checkpoint outperforming the domain-adapted LM, the two being practically equivalent, and the domain-adapted LM outperforming the official checkpoint, respectively. Following~\cite{bayesian-tutorial}, we treat a comparison as decisive when the highest of these posterior probabilities reaches $0.95$ and inconclusive otherwise. While a score delta conveys the magnitude and direction of a change, its corresponding highest PP conveys how consistently that change holds across tasks.

\section{Results}
\label{s:results}

We run every pre-training experiment following the recipe established in Subsection~\ref{ss:pretraining-details} and analyze how MLM and CLM losses behave. Since these losses are not comparable across experiments due to a variety of factors such as the loss definition itself, LM size, tokenizer vocabulary, among others, we care mostly about checking whether: (i) there is a big drop in the training and validation losses during the first stages of pre-training (i.e., the warm-up period and subsequent steps when the learning rate is close to its maximum value), then they continue dropping more conservatively; and (ii) there are no signs of overfitting represented by the gap between training and validation losses. We find that with our hyper-parameter selection these criteria are being met for all experiments to different degrees. The loss curves for all experiments can be found in the replication kit~\cite{replication-kit}.

Following Subsection~\ref{ss:evaluation}, we then fine-tune all LMs on the SELU and SuperGLUE tasks to subsequently compute the scores that enable us to evaluate domain adaptation and general-language understanding. We find that with our hyper-parameter selection at least 90\% and 85\% of fine-tuned LMs on the SELU and SuperGLUE tasks reach convergence before epoch 10; and 3\% and 20\%, respectively, have ill-defined confusion matrices. Particularly for SuperGLUE, these results confirm that, despite being released many years ago, it still poses a challenge for LMs at different scales. The absolute SELU and SuperGLUE scores can be found in Tables~\ref{tbl:selu-scores} and \ref{tbl:superglue-scores}, respectively, together with the score delta and the highest PP for relevant paired comparisons. Additional results per task can be found in the replication kit~\cite{replication-kit}.

Before diving into the analysis, we note that only a minority of the paired comparisons are decisive (highest PP $\geq 0.95$) and that decisiveness is markedly asymmetric between schemes. Under CPT, almost all comparisons are inconclusive on both benchmarks (i.e., only 2 SELU and 7 SuperGLUE comparisons reach the threshold), indicating that the changes CPT induces, whether gains or losses, are typically small and not consistent enough across tasks to be decisive. Under PTS, the picture reverses: most comparisons are decisive (16 on SELU, 13 on SuperGLUE), and every decisive PTS comparison favors the official checkpoint, reflecting large and consistent shortfalls relative to it. Since these probabilities are based on a Bayesian test, we can still interpret the probability of change as a trend, and of being equal as an indicator of stable non-difference.

\begin{table*}[ht]
\centering
\caption{Domain adaptation results in terms of SELU scores. Highest PP stands for the highest posterior probability from the Bayesian-signed rank test, where A corresponds to the official checkpoint and B the domain-adapted LM; the arrow indicates which relation attains this maximum ($\uparrow$ for $P(B>A)$, $\downarrow$ for $P(A>B)$, $\approx$ for $P(A=B)$). Cell shading encodes the magnitude and direction of $\Delta \text{SELU}$ (blue~=~gain, red~=~loss, intensity saturating at $|\Delta \text{SELU}|=10\%$); boldface marks decisive comparisons (highest PP~$\geq 0.95$). Non-bold posteriors ($< 0.95$) indicate a directional leaning only.}
\label{tbl:selu-scores}
\resizebox{\textwidth}{!}{%
\begin{tabular}{lc|ccc|ccc|ccc|ccc}
\toprule
\multicolumn{1}{c}{\multirow{2}{*}{\textbf{LM family}}} & \multirow{2}{*}{\begin{tabular}[c]{@{}c@{}}\textbf{Official}\\\textbf{checkpoint}\end{tabular}} & \multicolumn{6}{c|}{\textbf{CPT}} & \multicolumn{6}{c}{\textbf{PTS}} \\
\multicolumn{1}{c}{} &  & \begin{tabular}[c]{@{}c@{}}\textbf{3.3B}\\\textbf{tokens}\end{tabular} & {\boldmath$\Delta \textbf{SELU}$} & \textbf{Highest PP} & \begin{tabular}[c]{@{}c@{}}{\boldmath$6.01 \times e^{18}$}\\\textbf{FLOPS}\end{tabular} & {\boldmath$\Delta \textbf{SELU}$} & \textbf{Highest PP} & \begin{tabular}[c]{@{}c@{}}\textbf{3.3B}\\\textbf{tokens}\end{tabular} & {\boldmath$\Delta \textbf{SELU}$} & \textbf{Highest PP} & \begin{tabular}[c]{@{}c@{}}{\boldmath$6.01 \times e^{18}$}\\\textbf{FLOPS}\end{tabular} & {\boldmath$\Delta \textbf{SELU}$} & \textbf{Highest PP} \\ 
\hline
BERT base & \pct{0.7161} & \pct{0.7297} & \dcell{0.0136} & \pp{B>A}{0.747} & \pct{0.742} & \dcell{0.0259} & \pp{B>A}{0.998} & \pct{0.6522} & \dcell{-0.0639} & \pp{A>B}{0.9997} & \pct{0.6938} & \dcell{-0.0222} & \pp{A>B}{0.9946} \\
BERT large & \pct{0.6264} & \pct{0.6956} & \dcell{0.0691} & \pp{B>A}{0.7251} & \pct{0.6956} & \dcell{0.0691} & \pp{B>A}{0.7251} & \pct{0.2105} & \dcell{-0.416} & \pp{A>B}{1.0} & \pct{0.2105} & \dcell{-0.416} & \pp{A>B}{1.0} \\
RoBERTa base & \pct{0.7106} & \pct{0.7292} & \dcell{0.0185} & \pp{B>A}{0.8171} & \pct{0.7351} & \dcell{0.0244} & \pp{A=B}{0.5365} & \pct{0.7053} & \dcell{-0.0054} & \pp{A>B}{0.7842} & \pct{0.7329} & \dcell{0.0223} & \pp{A=B}{0.5182} \\
RoBERTa large & \pct{0.5738} & \pct{0.6677} & \dcell{0.0939} & \pp{B>A}{0.7967} & \pct{0.6677} & \dcell{0.0939} & \pp{B>A}{0.7967} & \pct{0.2083} & \dcell{-0.3655} & \pp{A>B}{0.9998} & \pct{0.2083} & \dcell{-0.3655} & \pp{A>B}{0.9998} \\
ModernBERT base & \pct{0.7093} & \pct{0.7346} & \dcell{0.0253} & \pp{B>A}{0.8332} & \pct{0.7307} & \dcell{0.0214} & \pp{B>A}{0.7226} & \pct{0.7123} & \dcell{0.003} & \pp{B>A}{0.5046} & \pct{0.7173} & \dcell{0.008} & \pp{B>A}{0.5477} \\
ModernBERT large & \pct{0.7241} & \pct{0.6899} & \dcell{-0.0342} & \pp{A>B}{0.6263} & \pct{0.694} & \dcell{-0.0301} & \pp{A>B}{0.8648} & \pct{0.7161} & \dcell{-0.008} & \pp{A>B}{0.5655} & \pct{0.7198} & \dcell{-0.0042} & \pp{A>B}{0.7472} \\
GPT-2 small & \pct{0.7338} & \pct{0.7379} & \dcell{0.0041} & \pp{A=B}{0.7975} & \pct{0.7467} & \dcell{0.0129} & \pp{A=B}{0.7888} & \pct{0.7165} & \dcell{-0.0173} & \pp{A>B}{0.7433} & \pct{0.7196} & \dcell{-0.0142} & \pp{A>B}{0.7594} \\
GPT-2 medium & \pct{0.7397} & \pct{0.7551} & \dcell{0.0153} & \pp{A=B}{0.6096} & \pct{0.7551} & \dcell{0.0153} & \pp{A=B}{0.6096} & \pct{0.7114} & \dcell{-0.0283} & \pp{A>B}{0.9562} & \pct{0.7114} & \dcell{-0.0283} & \pp{A>B}{0.9562} \\
GPT-2 large & \pct{0.7464} & \pct{0.7389} & \dcell{-0.0075} & \pp{A=B}{0.6152} & \pct{0.7504} & \dcell{0.0041} & \pp{B>A}{0.5165} & \pct{0.6868} & \dcell{-0.0595} & \pp{A>B}{0.9832} & \pct{0.6646} & \dcell{-0.0817} & \pp{A>B}{0.9994} \\
GPT-2 xl & \pct{0.7546} & \pct{0.7458} & \dcell{-0.0088} & \pp{A=B}{0.5823} & \pct{0.7461} & \dcell{-0.0085} & \pp{A=B}{0.7017} & \pct{0.7077} & \dcell{-0.0469} & \pp{A>B}{0.9997} & \pct{0.6425} & \dcell{-0.1121} & \pp{A>B}{1.0} \\
Llama 3.2 1B & \pct{0.7392} & \pct{0.7379} & \dcell{-0.0013} & \pp{A=B}{0.5394} & \pct{0.7478} & \dcell{0.0086} & \pp{B>A}{0.6671} & \pct{0.7006} & \dcell{-0.0387} & \pp{A>B}{0.9954} & \pct{0.6928} & \dcell{-0.0464} & \pp{A>B}{0.9959} \\
Llama 3.2 3B & \pct{0.7462} & \pct{0.7435} & \dcell{-0.0028} & \pp{A=B}{0.4286} & \pct{0.7524} & \dcell{0.0062} & \pp{B>A}{0.5051} & \pct{0.6654} & \dcell{-0.0808} & \pp{A>B}{1.0} & \pct{0.6479} & \dcell{-0.0983} & \pp{A>B}{1.0} \\
\hline
CodeBERT base & \pct{0.7138} & \pct{0.7147} & \dcell{0.0009} & \pp{A=B}{0.5037} & \pct{0.7353} & \dcell{0.0215} & \pp{B>A}{0.7372} &  &  &  &  &  &  \\
CodeLlama 7B & \pct{0.7135} & \pct{0.7376} & \dcell{0.0241} & \pp{B>A}{0.5935} & \pct{0.6886} & \dcell{-0.0249} & \pp{A>B}{0.7781} &  &  &  &  &  &  \\
StarCoder2 3B & \pct{0.7316} & \pct{0.7537} & \dcell{0.0221} & \pp{B>A}{0.865} & \pct{0.7602} & \dcell{0.0287} & \pp{B>A}{0.9545} &  &  &  &  &  &  \\
StarCoder2 7B & \pct{0.6797} & \pct{0.7525} & \dcell{0.0728} & \pp{B>A}{0.914} & \pct{0.7373} & \dcell{0.0575} & \pp{B>A}{0.7468} &  &  &  &  &  &  \\
\hline
min. & \pct{0.5738} & \pct{0.6677} & \dpp{-0.0342} &  & \pct{0.6677} & \dpp{-0.0301} &  & \pct{0.2083} & \dpp{-0.4160} &  & \pct{0.2083} & \dpp{-0.4160} &  \\
avg. & \pct{0.7099} & \pct{0.7290} & \dpp{0.0191} &  & \pct{0.7303} & \dpp{0.0204} &  & \pct{0.6161} & \dpp{-0.0939} &  & \pct{0.6135} & \dpp{-0.0966} &  \\
max. & \pct{0.7546} & \pct{0.7551} & \dpp{0.0939} &  & \pct{0.7602} & \dpp{0.0939} &  & \pct{0.7165} & \dpp{0.0030} &  & \pct{0.7329} & \dpp{0.0223} &  \\
\bottomrule
\end{tabular}%
}
\end{table*}

\begin{table*}[ht]
\centering
\caption{General-language results in terms of SuperGLUE scores. Reading is the same as for Table~\ref{tbl:selu-scores}.}
\label{tbl:superglue-scores}
\resizebox{\textwidth}{!}{%
\begin{tabular}{lc|ccc|ccc|ccc|ccc}
\toprule
\multicolumn{1}{c}{\multirow{2}{*}{\textbf{LM family}}} & \multirow{2}{*}{\begin{tabular}[c]{@{}c@{}}\textbf{Official}\\\textbf{checkpoint}\end{tabular}} & \multicolumn{6}{c|}{\textbf{CPT}} & \multicolumn{6}{c}{\textbf{PTS}} \\
\multicolumn{1}{c}{} &  & \begin{tabular}[c]{@{}c@{}}\textbf{3.3B}\\\textbf{tokens}\end{tabular} & {\boldmath$\Delta \textbf{SGLUE}$} & \textbf{Highest PP} & \begin{tabular}[c]{@{}c@{}}{\boldmath$6.01 \times e^{18}$}\\\textbf{FLOPS}\end{tabular} & {\boldmath$\Delta \textbf{SGLUE}$} & \textbf{Highest PP} & \begin{tabular}[c]{@{}c@{}}\textbf{3.3B}\\\textbf{tokens}\end{tabular} & {\boldmath$\Delta \textbf{SGLUE}$} & \textbf{Highest PP} & \begin{tabular}[c]{@{}c@{}}{\boldmath$6.01 \times e^{18}$}\\\textbf{FLOPS}\end{tabular} & {\boldmath$\Delta \textbf{SGLUE}$} & \textbf{Highest PP} \\ 
\hline
BERT base & \pct{0.5928} & \pct{0.5514} & \dcell{-0.0413} & \pp{A>B}{0.4508} & \pct{0.5766} & \dcell{-0.0162} & \pp{B>A}{0.7373} & \pct{0.486} & \dcell{-0.1068} & \pp{A>B}{0.7635} & \pct{0.4858} & \dcell{-0.1069} & \pp{A>B}{0.9087} \\
BERT large & \pct{0.5741} & \pct{0.5753} & \dcell{0.0012} & \pp{B>A}{0.4351} & \pct{0.5753} & \dcell{0.0012} & \pp{B>A}{0.4351} & \pct{0.4486} & \dcell{-0.1254} & \pp{A>B}{0.982} & \pct{0.4486} & \dcell{-0.1254} & \pp{A>B}{0.982} \\
RoBERTa base & \pct{0.6483} & \pct{0.6021} & \dcell{-0.0462} & \pp{A>B}{0.9985} & \pct{0.6362} & \dcell{-0.0121} & \pp{A>B}{0.9842} & \pct{0.5374} & \dcell{-0.1109} & \pp{A>B}{0.9984} & \pct{0.5364} & \dcell{-0.1119} & \pp{A>B}{0.9984} \\
RoBERTa large & \pct{0.6523} & \pct{0.6325} & \dcell{-0.0198} & \pp{B>A}{0.7029} & \pct{0.6325} & \dcell{-0.0198} & \pp{B>A}{0.7029} & \pct{0.4413} & \dcell{-0.211} & \pp{A>B}{0.9407} & \pct{0.4413} & \dcell{-0.211} & \pp{A>B}{0.9407} \\
ModernBERT base & \pct{0.572} & \pct{0.5475} & \dcell{-0.0245} & \pp{A>B}{0.8686} & \pct{0.5494} & \dcell{-0.0226} & \pp{A>B}{0.6914} & \pct{0.5644} & \dcell{-0.0076} & \pp{A>B}{0.66} & \pct{0.552} & \dcell{-0.02} & \pp{A>B}{0.8049} \\
ModernBERT large & \pct{0.6788} & \pct{0.623} & \dcell{-0.0558} & \pp{A>B}{0.9845} & \pct{0.6896} & \dcell{0.0108} & \pp{B>A}{0.5018} & \pct{0.561} & \dcell{-0.1178} & \pp{A>B}{0.9984} & \pct{0.5494} & \dcell{-0.1294} & \pp{A>B}{0.9983} \\
GPT-2 small & \pct{0.587} & \pct{0.588} & \dcell{0.001} & \pp{B>A}{0.7906} & \pct{0.5922} & \dcell{0.0052} & \pp{B>A}{0.83} & \pct{0.5646} & \dcell{-0.0225} & \pp{A>B}{0.7838} & \pct{0.5548} & \dcell{-0.0322} & \pp{A>B}{0.6069} \\
GPT-2 medium & \pct{0.5773} & \pct{0.6052} & \dcell{0.0279} & \pp{B>A}{0.9554} & \pct{0.6052} & \dcell{0.0279} & \pp{B>A}{0.9554} & \pct{0.5293} & \dcell{-0.048} & \pp{A>B}{0.9436} & \pct{0.5293} & \dcell{-0.048} & \pp{A>B}{0.9436} \\
GPT-2 large & \pct{0.6075} & \pct{0.6006} & \dcell{-0.0069} & \pp{A>B}{0.7821} & \pct{0.6218} & \dcell{0.0143} & \pp{A>B}{0.7491} & \pct{0.5423} & \dcell{-0.0651} & \pp{A>B}{0.9833} & \pct{0.5239} & \dcell{-0.0836} & \pp{A>B}{0.9914} \\
GPT-2 xl & \pct{0.5977} & \pct{0.6116} & \dcell{0.014} & \pp{B>A}{0.7395} & \pct{0.6339} & \dcell{0.0363} & \pp{B>A}{0.7608} & \pct{0.5582} & \dcell{-0.0395} & \pp{A>B}{0.9312} & \pct{0.541} & \dcell{-0.0567} & \pp{A>B}{0.9878} \\
Llama 3.2 1B & \pct{0.6677} & \pct{0.6798} & \dcell{0.0121} & \pp{A>B}{0.5292} & \pct{0.6647} & \dcell{-0.003} & \pp{B>A}{0.6836} & \pct{0.5541} & \dcell{-0.1136} & \pp{A>B}{0.9937} & \pct{0.5492} & \dcell{-0.1185} & \pp{A>B}{0.9904} \\
Llama 3.2 3B & \pct{0.7185} & \pct{0.7215} & \dcell{0.003} & \pp{B>A}{0.4166} & \pct{0.6793} & \dcell{-0.0392} & \pp{B>A}{0.5584} & \pct{0.5133} & \dcell{-0.2052} & \pp{A>B}{0.9968} & \pct{0.5146} & \dcell{-0.2039} & \pp{A>B}{0.9902} \\
\hline
CodeBERT base & \pct{0.612} & \pct{0.6309} & \dcell{0.0189} & \pp{B>A}{0.7035} & \pct{0.6118} & \dcell{-0.0002} & \pp{B>A}{0.8254} &  &  &  &  &  &  \\
CodeLlama 7B & \pct{0.6058} & \pct{0.6468} & \dcell{0.041} & \pp{A=B}{0.7} & \pct{0.6709} & \dcell{0.0651} & \pp{B>A}{0.5251} &  &  &  &  &  &  \\
StarCoder2 3B & \pct{0.5693} & \pct{0.6402} & \dcell{0.0709} & \pp{B>A}{0.9806} & \pct{0.6399} & \dcell{0.0706} & \pp{B>A}{0.9915} &  &  &  &  &  &  \\
StarCoder2 7B & \pct{0.6477} & \pct{0.6582} & \dcell{0.0105} & \pp{A>B}{0.4915} & \pct{0.6639} & \dcell{0.0162} & \pp{B>A}{0.4487} &  &  &  &  &  &  \\
\hline
min. & \pct{0.5693} & \pct{0.5475} & \dpp{-0.0558} &  & \pct{0.5494} & \dpp{-0.0392} &  & \pct{0.4413} & \dpp{-0.211} &  & \pct{0.4413} & \dpp{-0.211} &  \\
avg. & \pct{0.6193} & \pct{0.6197} & \dpp{0.0004} &  & \pct{0.6277} & \dpp{0.0084} &  & \pct{0.525} & \dpp{-0.0978} &  & \pct{0.5189} & \dpp{-0.104} &  \\
max. & \pct{0.7185} & \pct{0.7215} & \dpp{0.0709} &  & \pct{0.6896} & \dpp{0.0706} &  & \pct{0.5646} & \dpp{-0.0076} &  & \pct{0.5548} & \dpp{-0.02} &  \\
\bottomrule
\end{tabular}%
}
\end{table*}

\subsection{CPT effects on understanding of SE textual artifacts}

Relative to the official checkpoints, CPT improves domain adaptation only modestly and rarely decisively. Across budgets, only 2 of the highest PPs are $\geq 0.95$. Under the constant-token budget of 3.3B tokens, the average gain is larger for encoders ($\Delta \text{SELU} = 2.7\%$, avg.) than for decoders ($\Delta \text{SELU} = 1.3\%$, avg.). The weakest official checkpoints have the strongest movements: StarCoder2 7B and BERT large gain the most ($\Delta \text{SELU} = 7.3\%, 6.9\%$, resp.), and RoBERTa large posts the largest overall gain ($\Delta \text{SELU} = 9.4\%$) while still ranking lowest before and after CPT. The PPs greater than 0.725 indicate that while these differences are not decisive, it is likely that they are not random. At the other end, ModernBERT large is the only LM to lose ground under this budget ($\Delta \text{SELU} = -3.4\%$). The generalist decoders and CodeBERT base change by a negligible margin, indicating that there is no practical impact on domain adaptation.

Under the compute-matched budget of $6.01 \times e^{18}$ FLOPS, the pattern is similar, but the allocation of tokens by LM size becomes visible. For most of the smallest LMs (BERT base, RoBERTa base, GPT-2 small, and CodeBERT base), pre-training under this budget means that more tokens are seen in comparison to the  constant token budget. These higher numbers of tokens translate into larger domain adaptation gains and, for BERT base in particular, the results become decisive. Conversely, for the largest LMs (Llama 3.2 and StarCoder2 series), where the compute-matched budget represents far fewer domain-specific tokens, gains become tiny to moderate, the only decisive being StarCoder2 3B ($\Delta \text{SELU} = 2.9\%$). As under the constant-token budget, ModernBERT large loses ground ($\Delta \text{SELU} = -3.0\%$), joined by CodeLlama 7B ($\Delta \text{SELU} = -2.5\%$), both with relatively high PPs in favor of the official checkpoints.

\subsection{CPT effects on general-language retention}

The cost of CPT on general-language retention is relatively small on average but uneven; however, it does not represent a systematic forgetting pattern across all LMs. We observe clear differences between decoders and encoders. Decoder LMs show consistent gains under both budgets ($\Delta \text{SGLUE} = 2.0\%$, avg.), although with a few consistency issues to the point that for Llama 3.2 losses are arguable since their highest PPs lean toward the domain-adapted LMs outperforming the official checkpoints. Encoders are more exposed to general-language forgetting by losing ground in almost all cases, except for BERT large and CodeBERT base, which remain mostly stable. The losses for RoBERTa base and ModernBERT large are even among the few results that are decisive, but the latter deserves closer attention: adding only 13\% more domain-specific tokens, from 2.9B (compute-matched) to 3.3B (constant-token), flips its $\Delta \text{SGLUE}$ from a slight gain to a large loss, suggesting that general-language forgetting can happen relatively fast during domain adaptation.

\subsection{PTS for the understanding of SE textual artifacts}

Where CPT slightly improves domain adaptation ($\Delta \text{SELU} = 1.5\%$, avg., constant-token), PTS falls well short ($\Delta \text{SELU} = -9.4\%$, avg., constant-token), placing it about $11\%$ below CPT. This shortfall is decisive for most PTS LMs (16 of them across both budgets), in stark contrast to the mostly inconclusive changes of CPT. In other words, PTS clearly does not match CPT for domain adaptation at these budgets.

Still, the gap is governed by how many domain-specific tokens the budget affords and narrows wherever PTS is given more. For the smallest LMs (BERT base, RoBERTa base, ModernBERT base, and GPT-2 small), the compute-matched budget represents more tokens than the constant-token budget, and the PTS shortfall shrinks accordingly ($\Delta \text{SELU} = -2.1\% \rightarrow -0.15\%$, avg.). We observe a similar behavior for the larger LMs (GPT-2 large, xl, and the Llama 3.2 series), where the compute-matched budget supplies only a fraction of tokens, closing the gap by pre-training on more tokens ($\Delta \text{SELU} = -8.5\% \rightarrow -5.7\%$, avg.). ModernBERT large is the lone case in which PTS is not worse than CPT under either budget, but only because CPT itself already regresses the official checkpoint. The trends with respect to the relationship between constant-token and compute-matched budgets are consistent with the Chinchilla-optimal budget allocation framework~\cite{chinchilla}: lacking the initial advantage of a pre-trained checkpoint, PTS approaches CPT only when the budget is sufficiently token-rich.

\subsection{PTS for general-language emergence}

PTS does not reach the level of general-language understanding achieved by the official checkpoints and CPT, meaning that the differences against the official checkpoints are decisive or have the highest PPs above 0.607. Still, as the result for ModernBERT base suggests, it is possible to reach a competitive level in selected scenarios. This highlights the language richness and diversity within our corpus, which is dominated by SE textual artifacts from GitHub and Stack Overflow and has no exposure to broad general-domain text sources like Wikipedia or Common Crawl.

\subsection{Pre-training scheme across budgets, LM families, and sizes}

Taken together, the four comparisons answer our research question. By scheme, CPT is the safer and stronger choice: it yields small domain adaptation gains while largely preserving general-language understanding, whereas PTS pays a large and usually decisive penalty on both benchmarks relative to the official checkpoints. This asymmetry is the dominant signal in our results, with CPT comparisons mostly inconclusive rather than highlighting weak differences, and PTS comparisons reflecting mostly significant losses.

By LM family and size, encoders extract the largest domain adaptation gains from CPT at the cost of modest losses in general-language understanding. Decoders show little domain adaptation movement yet retain or slightly improve general-language understanding. For larger LMs, the CPT gains shrink and the PTS gaps widen under the compute-matched budget, whereas the smallest LMs benefit most from either scheme when the budget is token-rich.

Being fully aware of the difference between tokens and compute cost is necessary. This distinction was already raised by the seminal Chinchilla paper~\cite{chinchilla} for pre-training LMs with a GPT-3-like architecture from scratch. Our results demonstrate that the same consideration applies to domain adaptation: LMs require seeing a large number of tokens and, the larger the LM, the more tokens must be used. Thus, domain adaptation should only be considered for large LMs if sufficient domain-specific tokens and compute power are available.

\section{Discussion}
\label{s:discussion}

In general, our results have a clear red line: for adapting LMs to better understand SE textual artifacts, reusing an existing LM (as-is or via CPT) dominates training a domain-native one from scratch (PTS), as discussed in prior research~\cite{validity-se, bertoverflow}. We organize the discussion around what that reuse actually buys, how the picture varies by LM architecture and size, whether the existing LM was pre-trained on code, why a budget-aware comparison protocol is necessary to observe these effects, and what all of this implies for SE researchers and practitioners.

\subsection{Reuse over rebuild}

The dominant signal in our results is an asymmetry: CPT yields small and mostly inconclusive gains, whereas PTS pays large and usually-decisive losses on both domain adaptation and general-language understanding. Reading against official checkpoints that anchor every comparison reframes the practical choice as three-tiered rather than only CPT vs. PTS. The official checkpoint, used as-is, is the cheapest option and already a strong one: CPT improves domain adaptation only modestly and rarely decisively, while leaving general-language understanding essentially unchanged on average, notably without any replay or regularization to counter forgetting, which suggests that CPT on SE text may not require dedicated anti-forgetting techniques at these budgets. CPT is therefore best understood as an optional, low-risk refinement on top of the official checkpoint rather than a mandatory step. It helps most where the existing LM is weak and rarely hurts.

A plausible explanation for these muted CPT gains is the limited headroom. Our corpus is drawn from GitHub, Stack Overflow, Jira, and arXiv, precisely the kind of public text that pervades the web-scale corpora on which modern LMs may have been originally pre-trained. Where domain-adaptive pre-training targets text that is genuinely out-of-distribution, such as finance, the reported gains are larger~\cite{finpythia,domain-adaptation1}; SE text, by contrast, is only weakly out-of-domain for these LMs, so CPT largely reinforces what they already encoded rather than introducing new knowledge.

We emphasize that this is a decision about starting points for fine-tuning, not about prompting. The LMs we study are encoders and decoders, and every SELU and SuperGLUE score is obtained after task-specific fine-tuning; none is used zero-shot. The operative question for a practitioner is thus whether to do a CPT pass before fine-tuning an existing LM and, on our evidence, the burden of proof falls on CPT to justify its added compute.

\subsection{Latest models might not need adaptation}

Another relevant aspect from our results is that some of the latest and larger LMs we study (ModernBERT, Llama 3.2) are mostly hurt or have only negligible gains from CPT. This points to a consideration that deserves further analysis: later training regimes that have more careful data schedules~\cite{data-mixtures1,data-mixtures2,data-mixtures3} may benefit less, if at all, from pre-training on SE text. Further research should try to establish if this is an artifact of the mixture data and LMs in our setting or a general trend.

\subsection{Code-focused LMs and the case for integrated code and text understanding}

A more surprising pattern concerns the code-focused LMs, and it connects directly to the motivation of this study: code-focused LMs may not be optimized for understanding of SE textual artifacts. Because they start out poor in natural language, they carry the most headroom, and CPT on our purely-text corpus lifts them on both axes with no apparent trade-off for language-understanding. We note that our study does not control for losses in code-focused tasks. StarCoder2 7B posts the largest decoder SELU gain, StarCoder2 3B is among the few decisive SELU gains, and together with CodeLlama 7B, shows the largest SuperGLUE gains. Adapting a code-focused LM on SE text adds general-language understanding rather than eroding it, precisely because its code-heavy pre-training under-weighted natural language from the beginning. This joint training of SE texts and code should, therefore, be further explored to better understand the interaction between both modes and, ideally, achieve small and efficient models capable in both tasks. We highlight that StarCoder2 already used a carefully designed mixture of code and textual artifacts, but that the code was the majority of the training. Other mixtures might lead to better overall results, warranting the study of specific training curricula for the training of LMs for both code and natural language SE tasks.

\subsection{Guidance for SE researchers and practitioners}

Taken together, our findings translate into concrete guidance: Default to reusing an official checkpoint and add CPT when there is a specific reason to expect a gain, namely a weak SE text baseline, a code-focused LM targeting mixed use (code + text), or a small LM under a token-rich budget. Reserve PTS for settings where no suitable checkpoint exists or domain-native representations are themselves the object of study. Always consider both the token and compute budget together. Finally, evaluate on both SE-specific and general-language benchmarks: our results show that conclusions drawn from either axis alone or from an uncontrolled budget can be misleading.

\section{Threats to Validity}
\label{s:threats}

In this section, we systematically examine potential threats to the validity of our work in four dimensions to provide context for interpretation of the results. By identifying these threats, we aim to guide future work towards improving the robustness and generalizability of LMs in SE.

\subsection{Construct Validity}

Our two constructs, domain adaptation and general-language retention, are operationalized through a single benchmark each, so our measurements are only as faithful as those proxies. However, we note that SELU~\cite{selu} is, to the best of our knowledge, the broadest available testbed for SE textual artifacts. Similarly, SuperGLUE~\cite{superglue} was designed to be solvable without domain-specific knowledge, which makes it a reasonable but indirect proxy for general-language capabilities. Because we collapse each benchmark into a single averaged score, per-task effects, including cases where adaptation helps some tasks and hurts others, are hidden by construction; per task results are available in the replication kit~\cite{replication-kit}. 

\subsection{Internal Validity}

Pre-training is known to be unstable, and several of our design choices could confound the comparison between schemes. We run a single global seed (42) for both pre-training and fine-tuning, so we cannot separate genuine effects from run-to-run variance. Such comparisons are not possible with our available compute resources. The ModernBERT large drop under CPT is a concrete case where we cannot rule out seed or hyper-parameter sensitivity as the explanation. CPT reuses the tokenizer published with each official checkpoint, whereas PTS trains a new tokenizer on our SE corpus matching the algorithm and vocabulary size. Part of the CPT vs. PTS gap may therefore reflect differences in tokenization rather than the pre-training scheme alone. Our study deliberately prioritizes comparability under constant-token and compute-matched budgets over per-experiment optimization, so the shared hyper-parameters (global batch size, maximum learning rate of $1 \times e^{-5}$, weight decay, cosine schedule) are a compromise rather than an optimum for any single LM. Moreover, the need to switch some fine-tuning runs to a maximum learning rate of $1 \times e^{-4}$ for smaller datasets illustrates that no single configuration is best everywhere. Lastly, our convergence and ill-defined confusion-matrix checks are met for most but not all experiments (at least 90\%/85\% convergence on SELU/SuperGLUE), so a minority of fine-tuning runs may under-report achievable performance.

A specific internal validity concern is that information leakage during the pre-training of models. While it is unlikely that SuperGLUE data is part of our GitHub and Stack Overflow-dominated corpus and while we specifically cleaned our corpus to avoid duplication with SELU, we cannot be certain whether the training of official checkpoints included data from these tasks. In such cases, the performance of official checkpoints is possibly over-estimated. 

\subsection{External Validity}

Our corpus is assembled from four large, public sources (GitHub, Stack Overflow, Jira, arXiv), so our findings may not transfer to proprietary or industrial SE artifacts, to closed issue trackers, or to languages other than English, which we enforce through a language filter. Because we mask code blocks and focus the study on SE text, our results speak to natural-language understanding of SE textual artifacts and should not be read as claims about code-focused adaptation. We cover a range of encoder and decoder families from 108M to 7.2B parameters, but this is still a limited slice of the LM space (very large models, impact of post-training like RLHF~\cite{rlhf} or DPO~\cite{dpo}, etc.).

\subsection{Conclusion Validity}

Our budget-aware protocol rests on the FLOPS approximation according to Equation~\ref{eq:flops-vs-tokens} that, although it was derived primarily for decoders, we apply uniformly to encoders and ignore architecture and family-specific factors (e.g., local/global attention in ModernBERT). Recent MLM scaling research suggests that encoders follow different compute-optimal ratios~\cite{compute-optimal-encoders}, so the compute-matched budget should be read as a principled control rather than an exact equalization. We sample only two budget points (at 3.3B tokens and $6.01 \times e^{18}$ FLOPS), limiting how confidently we can describe the shape of the budget-performance relationship.

Finally, statistical inference is similarly constrained. Running a single global seed, we cannot estimate run-to-run variance. The Bayesian signed-rank test rests on its own assumptions, i.e., it operates  on per task score differences against a ROPE of $\pm 0.1 \cdot d$ and its resolving power is bounded by the number of paired tasks (20 for SELU and 7 for SuperGLUE). With so few paired tasks, only large and consistent differences reach the $0.95$ threshold, which is why the small effects of CPT seldom become significant, while the large PTS shortfalls do.

\section{Conclusions}
\label{s:conclusions}


Our study shows that existing LMs are already a good starting point to fine-tune for understanding of SE textual artifacts, but that CPT can be used as a technique to slightly enhance LMs in this regard. The expected gains from CPT are larger for older encoder LMs at the cost of some loss on general-language understanding. PTS is inferior to starting from an existing LM. Future work should explore whether such pre-training can also lift code-focused tasks; our results already show that SE text improves general-language understanding of code-focused models. A combined training with code and SE text may help LMs to become better at tasks that require understanding both types of content, e.g., documentation or generation of code from comments.

Our SE corpus, pre-trained LMs, tokenizers, scripts, hyper-parameter configurations, and detailed results can be found in our replication kit~\cite{replication-kit}.



\bibliographystyle{IEEEtran}
\bibliography{refs}

\end{document}

%% file: preamble.tex
\documentclass[conference]{IEEEtran}

\IEEEoverridecommandlockouts

\usepackage{cite}
\usepackage{amsmath,amssymb,amsfonts}
\usepackage{algorithmic}
\usepackage{graphicx}
\usepackage{textcomp}
\usepackage{xcolor}
\def\BibTeX{{\rm B\kern-.05em{\sc i\kern-.025em b}\kern-.08em
    T\kern-.1667em\lower.7ex\hbox{E}\kern-.125emX}}

\usepackage{orcidlink}
\usepackage{multirow}
\usepackage{pifont}
\usepackage{colortbl}
\usepackage{arydshln}
\usepackage{booktabs}
\usepackage{xfp}
\usepackage{siunitx}
\usepackage{makecell}

\newcommand{\pctnum}[1]{\num[round-mode=places,round-precision=1]{\fpeval{#1*100}}}
\newcommand{\pct}[1]{\pctnum{#1}\%}
\newcommand{\dpp}[1]{\pctnum{#1}\%}

\newcommand{\dcell}[1]{%
  \xdef\dcellpct{\fpeval{min(round(abs(#1)/0.1*40),40)}}%
  \ifnum\fpeval{#1<0}=1\relax
    \cellcolor{red!\dcellpct}%
  \else
    \cellcolor{blue!\dcellpct}%
  \fi
  \dpp{#1}}
\newcommand{\ppnum}[1]{\num[round-mode=figures,round-precision=3,detect-weight=true,detect-family=true]{#1}}
\ExplSyntaxOn
\NewDocumentCommand{\ppdir}{m}{%
  \str_case:nnF {#1}
    { {B>A}{$\uparrow$} {A>B}{$\downarrow$} {A=B}{$\approx$} }
    {$P(#1)$} }
\NewDocumentCommand{\pp}{mm}{%
  \fp_compare:nNnTF {#2} < {0.95}
    { \ppdir{#1}~\ppnum{#2} }
    { \ppdir{#1}~\textbf{\ppnum{#2}} } }
\ExplSyntaxOff